\documentclass[10pt,conference]{IEEEtran}
\IEEEoverridecommandlockouts
\usepackage{cite}
\usepackage{amsmath,amssymb,amsfonts}
\usepackage{algorithmic}
\usepackage{graphicx}
\usepackage{textcomp}
\usepackage{xcolor}
\usepackage{graphicx}
\usepackage{caption}
\usepackage{subcaption}
\usepackage[utf8]{inputenc}
\usepackage{enumerate}
\usepackage{enumitem}
\usepackage{xcolor}
\usepackage{fbox}
\usepackage{diagbox}
\usepackage{listings}
\usepackage{lipsum}
\usepackage{listings}
\usepackage{hyperref}
\usepackage{amsmath}
\usepackage{cleveref}
\usepackage{dirtree} 
\usepackage{multirow}
\usepackage{multicol}
\usepackage[ruled,linesnumbered]{algorithm2e}
\usepackage{ulem}
\usepackage[utf8]{inputenc}
\usepackage{bbding}
\usepackage{wasysym}
\usepackage[flushleft]{threeparttable}
\usepackage{xurl}
\usepackage{pifont}
\usepackage{balance}
\usepackage{soul}
\usepackage{setspace}
\usepackage[breakable,skins]{tcolorbox}

\crefformat{section}{\S#2#1#3}
\crefname{figure}{Figure}{Figures}
\crefname{table}{Table}{Tables}
\crefname{algocf}{Algorithm.}{Algorithms.}
\Crefname{algocf}{Algorithm}{Algorithms}

\SetKwInput{kwInit}{Require}
\SetKwComment{Comment}{/* }{ */}

\newtcolorbox{boxA}{
    boxrule = 1pt,
    colframe = black 
}

\newcommand{\PeerSpin}{PeerSpin\xspace}
\newcommand{\PeerChecker}{PeerChecker\xspace}
\newcommand{\Tech}{Node-Replacement-Conflict based \PeerSpin Detection\xspace}
\newcommand{\code}[1]{\texttt{{\small \detokenize{#1}}}}

\def\BibTeX{{\rm B\kern-.05em{\sc i\kern-.025em b}\kern-.08em
    T\kern-.1667em\lower.7ex\hbox{E}\kern-.125emX}}
\begin{document}

\title{Understanding and Detecting Peer Dependency Resolving Loop in npm Ecosystem}

\author{
\IEEEauthorblockN{Xingyu Wang}
\IEEEauthorblockA{
    Zhejiang University \\
    Hangzhou, China \\
    wangxingyu@zju.edu.cn}
\and
\IEEEauthorblockN{Mingsen Wang}
\IEEEauthorblockA{
    Zhejiang University\\
    Hangzhou, China \\
    udehbsmw@zju.edu.cn}
\and
\IEEEauthorblockN{Wenbo Shen\IEEEauthorrefmark{1}}
\IEEEauthorblockA{
    Zhejiang University \\
    Hangzhou, China \\
    shenwenbo@zju.edu.cn}
\and
\IEEEauthorblockN{Rui Chang}
\IEEEauthorblockA{
    Zhejiang University \\
    Hangzhou, China \\
    crix1021@zju.edu.cn}
\thanks{\IEEEauthorrefmark{1}Corresponding author.}
}

\maketitle

\thispagestyle{plain}
\pagestyle{plain}

\begin{abstract}
As the default package manager for Node.js, \textit{npm} has become one of the largest package management systems in the world.
To facilitate dependency management for developers, npm supports a special type of dependency, \textit{Peer Dependency}, whose installation and usage differ from regular dependencies. 
However, conflicts between peer dependencies can trap the npm client into infinite loops, leading to resource exhaustion and system crashes. We name this problem \textit{\PeerSpin}.
Although \PeerSpin poses a severe risk to ecosystems, it was overlooked by previous studies, and its impacts have not been explored.

To bridge this gap, this paper conducts the first in-depth
study to understand and detect \PeerSpin in the npm
ecosystem.
First, by systematically analyzing the npm dependency resolution, we identify the root cause of \PeerSpin and characterize two peer dependency patterns to guide detection.
Second, we propose a novel technique called \textit{\Tech}, which leverages the state of the directory tree during dependency resolution to achieve accurate and efficient \PeerSpin detection.
Based on this technique, we developed a tool called \textit{\PeerChecker} to detect \PeerSpin.
Finally, we apply \PeerChecker to the entire npm ecosystem and find that 5,662 packages, totaling 72,968 versions, suffer from \PeerSpin. 
Until now, we have selected 100 problematic packages to report and received 28 confirmations.
We also open source all \PeerSpin analysis implementations, tools, and data sets to the public to help the community detect \PeerSpin issues and enhance the reliability of the npm ecosystem.
\end{abstract}

\section{Introduction}
With the widespread adoption of JavaScript and Node.js, \textit{npm}~\cite{NPMWebsite}, the default package management system for Node.js, has emerged as one of the largest and most influential open-source software ecosystems. 
By October 2023, the npm ecosystem had indexed over 2.5 million packages and over 36 million package versions, providing comprehensive metadata that includes version information and dependencies on other packages. 
This extensive repository supports various applications and services, enabling developers to share, discover, and manage code efficiently.

In npm, besides the regular dependency, it also introduced \textit{peer dependency} for a package to specify that it relies on a specific version of another package, which is expected to be installed higher in the dependency tree~\cite{PeerDepDefinition}.
Peer dependencies are usually used when a package is meant to be used alongside one specific package.
Specifically, \code{peerDependencies} are used to express the relationship between a plugin package and its host packages.
One example is \code{react-dom}~\cite{react-dom}, a plugin package of the host package \code{react}~\cite{react}. 
While \code{react-dom} itself does not depend on \code{react}, developers must ensure that the runtime environment has a compatible version of \code{react} before using \code{react-dom} properly. As a result, \code{react} is specified as a peer dependency of \code{react-dom}.
Peer dependency is critical to npm.
Our analysis shows that among the 2.5 million packages in npm, 61.15\% use peer dependencies, and over 40\% of newly released packages have contained peer dependencies in the last four years.

Though peer dependency is widely used in npm, its resolution is a mess due to historical reasons.
Early versions of npm (v3 to v6) don't resolve peer dependencies automatically. When detecting a peer dependency requirement, they only give warnings to remind developers to install these peer dependencies manually.
Since npm v7, the npm client has started to resolve peer dependencies automatically.
However, specific peer dependencies can trap the npm client in an infinite loop, which blocks and crashes the host PC system, causing dozens of reported issues~\cite{Issues}.
Most discussions on these issues focus on walking around the problems rather than addressing their root causes. As a result, the underlying reasons for infinite loops remain unexplored.

We named this infinite looping problem caused by peer dependency resolution as \textit{\PeerSpin}.
\PeerSpin can propagate through dependency chains. If a package with \PeerSpin issue is relied upon by downstream packages, the resolution of the downstream package may also be blocked or crashed by \PeerSpin.
Even worse, \PeerSpin is very hard to debug as the npm client gets stuck without giving any information on stuck points or stuck reasons.
Given the complexity of dependencies of npm projects, \PeerSpin poses a significant risk to the usability of the npm ecosystem.

Previous studies on the npm ecosystem have primarily focused on regular dependencies, such as vulnerability propagation~\cite{VulviaDepTree, SecImpact, wang2023plumber, 10.1145/3571848, zimmermann2019small}.
However, these studies do not consider the peer dependencies.
Specifically, Abate et al.~\cite{9054837} highlight the complexity of dependency resolution but do not analyze the additional complexity introduced by different types of dependencies.
Duan et al.~\cite{duan2020towards} provide a matrix of common security risks in package management systems for interpreted languages and conduct an ecosystem analysis.
Wyss et al.~\cite{wyss2022wolf} propose a privilege management system to counter installation-time attacks propagating through software dependency chains.
However, these studies do not address \PeerSpin or methods to detect \PeerSpin.
To the best of our knowledge, \PeerSpin has not been systematically studied so far.

To bridge this gap, this paper conducts the first in-depth study to understand and detect \PeerSpin in the whole npm ecosystem.
We begin by systematically investigating the dependency resolution process to uncover the root cause of \PeerSpin and propose two dependency patterns that could lead to it.
Following that, we design and implement an accurate and efficient detector, allowing us to quantify the \PeerSpin impacts on an ecosystem scale.
Though conceptually simple, we face two major challenges to achieving the analysis.

First, \PeerSpin issues are inherently complex to identify and understand. 
It is caused by a logical flaw in the dependency resolution algorithm associated with the directory tree state during installation.
The dynamic complexity of the directory tree state may prevent detection through static analysis, and \PeerSpin might not manifest until the dependency resolution process is executed, making it difficult to predict or identify in advance.
Meanwhile, \PeerSpin is highly related to the directory tree, and the tracking state of the tree further complicates detection efforts.
In addition, \PeerSpin causes the npm client to hang without providing helpful information. Even with all the installation logs, the complex dependencies make it difficult for developers or users to identify the stuck point.
Second, detecting \PeerSpin on an ecosystem scale is efficiently and accurately challenging.
Although long installation times may indicate \PeerSpin, relying solely on installation-time metrics for detection is inaccurate. 
The npm ecosystem is characterized by complex dependency trees, with each package typically depending on an average of 80 other packages~\cite{zimmermann2019small}.
This extensive dependency means the installation process can be time-consuming, even without \PeerSpin.
Additionally, long installation times can result from various benign factors, such as large package sizes, network latency, or the processing of numerous dependencies.
Consequently, using installation time as an indicator for detecting \PeerSpin can result in inefficiencies and inaccuracies.

To address these challenges, we examine the resolution algorithm and find that \PeerSpin occurs due to node replacement in the directory tree, forming a duplicate sequence of nodes placed in the tree during installation. 
We further propose a novel technique named \textit{\Tech} for detecting \PeerSpin automatically. 
We design and implement a tool, \textit{\PeerChecker}, which effectively addresses dependency resolution and integrates with detection methods to detect \PeerSpin for the entire npm ecosystem.

We conduct extensive evaluations to examine the accuracy and performance of \PeerChecker.
For accuracy, we compare the results of \PeerChecker of more than 80,000 packages with the behaviors of actual installation using the npm client, finding that the number of false positives and false negatives is zero, which illustrates the accuracy of our detection method.

Regarding performance, with 1,000 randomly selected package versions, \PeerChecker is 14 times faster than the npm client in dependency resolution, which allows us to extend our detection to the entire ecosystem scale.

To quantitatively analyze the impact of \PeerSpin issues, we ran \PeerChecker on the npm ecosystem, analyzing data up to October 2023. 
\PeerChecker detected 72,968 versions of 5,662 packages that could cause the \PeerSpin. We reported the problematic versions of 100 of these packages to their developers and received 28 confirmations.

In summary, our work makes four major contributions.
\begin{itemize}
    \item \textbf{New Study.} We conducted the first study of \PeerSpin issues in the npm ecosystem. Our findings help stakeholders understand the characteristics of \PeerSpin issues and provide guidance to fix these issues.
    
    \item \textbf{New Technique and Tool.} We propose a novel technique, \Tech, to achieve accurate \PeerSpin detection. We also design and implement a tool, \PeerChecker, to detect \PeerSpin for the entire npm ecosystem. 
    
    \item \textbf{Ecosystem Analysis.} We conduct the first ecosystem-scale analysis to reveal the \PeerSpin impacts. Our analysis shows that 72,968 versions of 5,662 packages suffer from \PeerSpin.  We selected 100 problematic packages and reported them to their developers, and eventually received 28 confirmations.

    \item \textbf{Community Contributions.} We open all \PeerSpin analysis tools and data sets to the public at \url{https://github.com/ZJU-SEC/PeerChecker}.
\end{itemize}

\begin{figure}[t]
    \centering
    \includegraphics[width=0.9\linewidth]{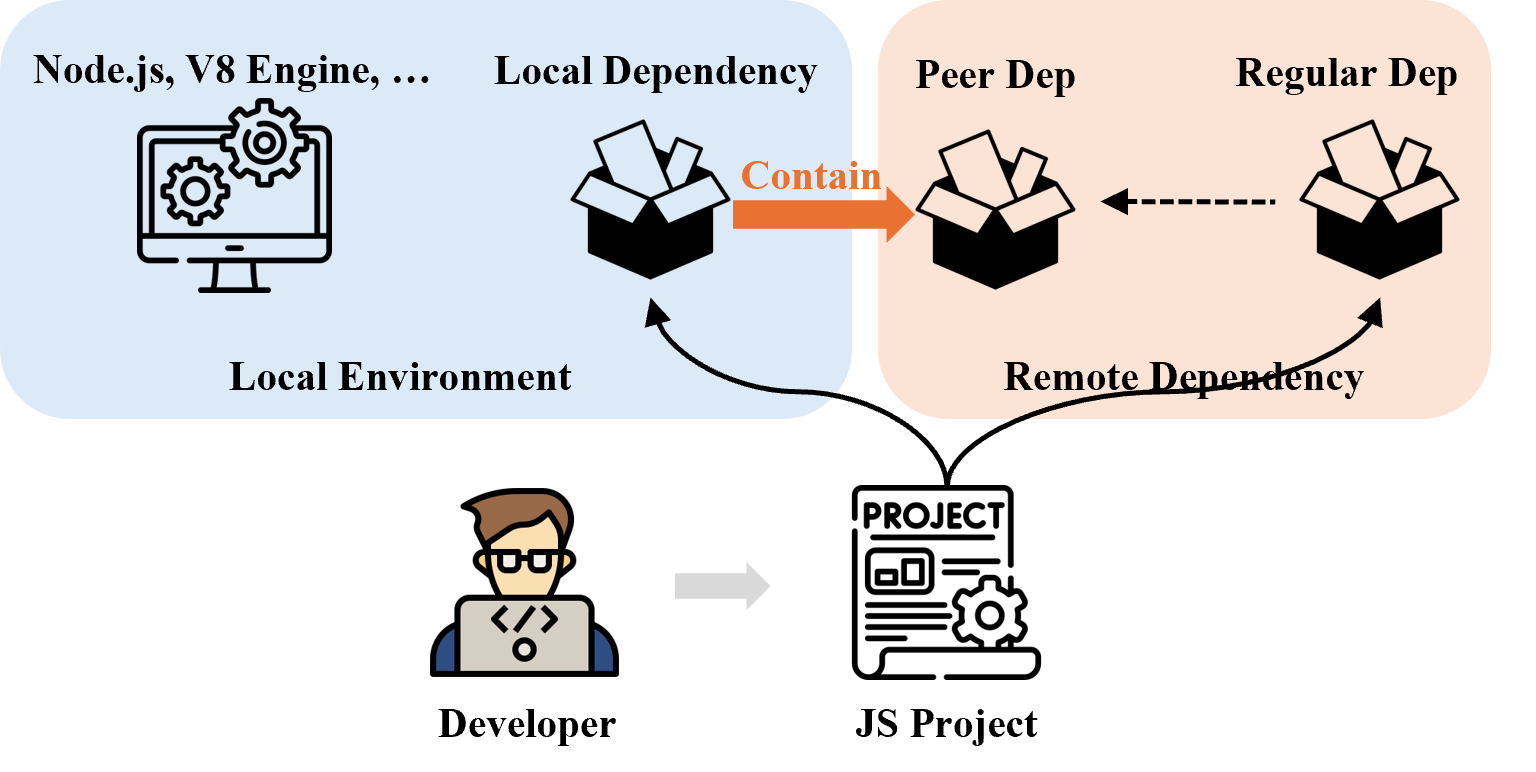}
    \caption{Peer dependencies of an npm package.}
    \label{fig:PeerIntro}
\end{figure}

\section{Background and Motivation Example}
\label{sec:background}

In this section, we first give preliminary knowledge of this paper. Then, we present a real-world example of \PeerSpin to show the motivation of this paper.

\subsection{Background}
\subsubsection{Peer Dependencies}

Peer dependencies specify that a package requires a particular version of another package to be provided by the consuming project, ensuring compatibility and preventing multiple instances of the dependency.
In \cref{fig:PeerIntro}, the developer introduces a third-party package as a regular dependency. This third-party package specifies its peer dependency. Consequently, for proper functionality, the developer must ensure that this peer dependency is satisfied in the local environment.

Before npm v7, the npm client did not automatically install sibling dependencies. The installation of peer dependencies relied on manual installation by developers. Since npm v7, the npm client has automatically installed peer dependencies.
However, due to the large number of dependencies in npm packages and the complexity of peer dependency rules, the npm client struggles to handle peer dependencies in certain dependency patterns.

There is some terminology that needs to be clarified.
\begin{itemize}[leftmargin=*]
    \item \textbf{PeerSet}: For a given package, the set of its direct and indirect peer dependencies forms its  \textit{PeerSet}. For example, in ~\cref{fig:PeerSpin}, \code{antd} and its peer dependencies, \code{react-dom} and \code{react}, form a \textit{PeerSet}.

    \vspace{5pt}
    \item \textbf{PeerSource and PeerEntry}: If a \textit{PeerSet} is introduced by a regular dependency, for all packages in the set,  the source package of this regular dependency is called \textit{PeerSource}, and the target package is called \textit{PeerEntry}.
    For example, in ~\cref{fig:PeerSpin}, \code{xydesign} is a \textit{PeerSource} for \code{react-dom} and \code{react}, and \code{antd} is their \textit{PeerEntry}.
    
\end{itemize}

\subsubsection{Dependency Loading}
\label{DependencyLoading}
In Node.js, the \code{require} function is used to load dependencies. When executing \code{require}, the function searches the installation directory structure.
It traverses the parent-level directories (containing the current directory) from the lowest to the highest level up to the root directory, attempting to find the target dependencies under \code{node_modules} in each parent directory.
Based on the search rules, npm's dependency management is essentially the organization of the installation directory tree, which is structured according to the dependency relationships between packages.

By the definition of peer dependency, in the node tree, all packages in a PeerSet should be able to be found by their PeerSource according to the loading rules.

\subsubsection{Dependency Model}
To ensure dependencies are loaded correctly, npm considers both the tree structure of installation directories and the graph structure of dependencies.
As a result, the npm dependency model combines both tree and graph structures, clearly distinguishing the relationships between nodes in these two structures. The key concepts of the npm dependency model are outlined below:

\vspace{5pt}
\noindent\textbf{Node and Edge}: Each node represents a specific package version, denoted as $nv$, and the edge represents the dependency relationship between packages, denoted as $<n_{from}, n_{to}, req, s>$ which means the package $n_{from}$ depends on package $n_{to}$, $req$ represents the version requirement from $n_{from}$ to $n_{to}$ and $s$ indicates the status of the edge. During the package installation process, the status of an edge can be classified as either \code{valid} or \code{invalid}. An edge is valid only if $n_{to}$ has at least one version that satisfies $n_{from}$. 

\vspace{5pt}
\noindent\textbf{Node Graph}: A graph composed of nodes and edges. The node graph represents the dependencies described in the package meta-information but does not represent the software installation directory structure.

\vspace{5pt}
\noindent\textbf{Node Tree}: When npm installs a package, it creates a directory structure to manage dependencies and facilitate efficient installation, updating, and usage of packages. This directory structure is referred to as a node tree. The \code{Root} represents the current package installation directory, and the tree represents the directory structure relationship of each package under the current installation directory. 

\begin{figure}[t]
    \centering
    \includegraphics[width=0.75\linewidth]{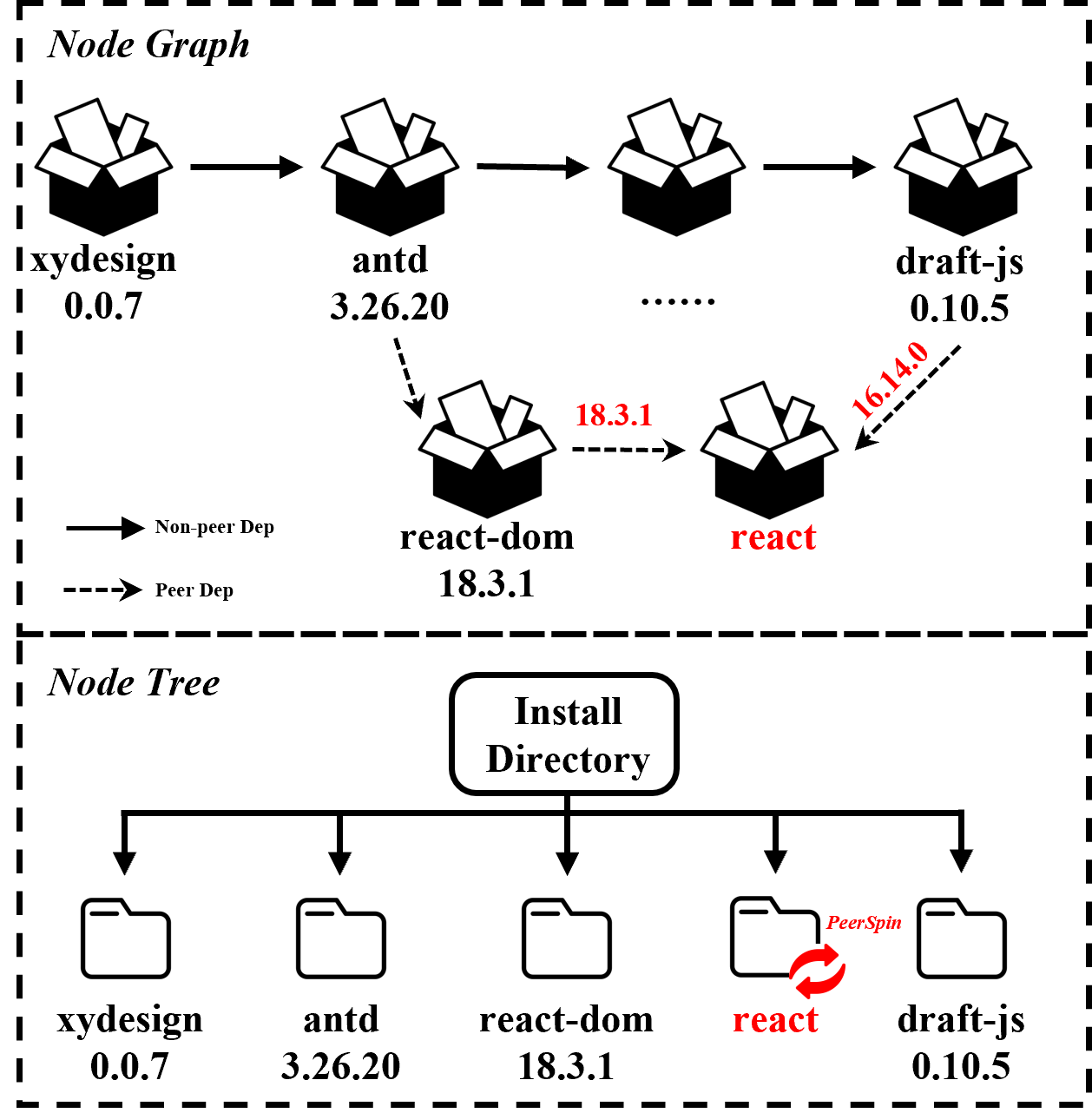}
    \caption{Motivating examples of a \PeerSpin issue in installing \code{xydesign}.}
    \label{fig:PeerSpin}
\end{figure}

The relative hierarchies of packages in the node tree and node graph may differ. 
As shown in \cref{fig:PeerSpin}, \code{react} is dependent on \code{react-dom}, and \code{draft-js}.
When installing \code{xydesign}, the npm client can place \code{react} at the same directory level as \code{react-dom} and \code{draft-js} to reduce redundancy.

\subsection{Motivation Example}
\Cref{fig:PeerSpin} gives a real-world example of \PeerSpin. The \textit{Node Graph} indicates the dependency requirement between packages in which \code{react} is introduced as a peer dependency through \code{draft-js} (directly) and \code{antd} (indirectly). 
When installing \code{xydesign}, the npm client creates a directory tree called \textit{Node Tree}. The structure of the node tree is optimized to reduce redundant installation of dependencies. 
In this example, all dependency packages are attempted to be installed at the same directory level.
However, \code{draft-js} and \code{antd} have different and incompatible version requirements for \code{react}, and they specify it as \code{peerDependencies}, which prevents \code{react} from being installed in a subdirectory under \code{draft-js} or \code{antd}.
The npm client will try to determine the exact version of \code{react}, but whichever version is chosen will inevitably fail to meet the version requirements of either \code{react-dom} or \code{draft-js}.
The npm client falls into an infinite loop where the two versions of \code{react} keep replacing each other, resulting in an interminable installation process, which we named \PeerSpin. 
The npm client can no longer respond to any user requests, and the resources it consumes continue to increase, eventually leading to a device crash.

Note that \PeerSpin is distinct from dependency conflicts.
If \code{react} is a regular dependency, the resolving algorithm can place the conflicting version within the \code{react-dom} or \code{draft-js} directory to remediate the conflict\cite{DepHell}. 
However, in this case, according to the peer dependency rules, \code{react} must be placed at the same level as its downstream packages (\code{react-dom} and \code{draft-js}).
\section{PeerSpin: Infinite Loop in Peer Dependency Resolving}
\label{PeerSpin}

To conduct \PeerSpin impact study on the npm ecosystem, we first need to understand how \PeerSpin occurs.
Therefore, we present the systematic study of the \PeerSpin issues in this section.
We explored the following two research questions:
\begin{itemize}
    \item \textbf{RQ1 (Root Cause)}: What is the root cause that the leading installation process cannot be terminated? 
    \item \textbf{RQ2 (Dependency Patterns)}: What are the common peer dependency patterns that can cause \PeerSpin issues?
\end{itemize}

\subsection{RQ1: Root Cause}
In earlier versions of npm (v3 to v6), peer dependencies were not automatically installed, the developer only received a warning that the peer dependencies were not installed.
Since npm v7, the npm client has introduced the \textit{Arborist} module (\code{@npmcli/arborist})~\cite{arborist} for dependency management.
The \textit{Arborist} module automatically installs peer dependencies~\cite{PeerAuto}, leading to the emergence of \PeerSpin issues.
Nodes and edges, node tree, node graph, and other elements of the npm dependency model are constructed and processed within the \textit{Arborist} module.
As the npm client has undergone numerous updates and feature additions, the specification document has become incomplete and outdated.
It does not accurately reflect the dependency resolution.
This paper examines the dependency resolution process by modeling and debugging the source code of the \textit{Arborist} module.
We then uncover that the dependency resolution algorithm of the \textit{Arborist} module still has logical flaws and inadequate handling of special cases, leading to the \PeerSpin problem in peer dependency resolution identified in this paper.

Our analysis shows that the npm dependency resolution algorithm uses breadth-first traversal.
The algorithm maintains a tree and a queue of pending nodes. 
Nodes in the queue are already in the tree, but their dependencies may not be satisfied.
The algorithm removes these nodes from the queue, adds new nodes to the tree to satisfy dependencies, and places the new nodes back in the queue.
Eventually, the edges of each node in the tree are valid.
The core resolving process involves three sub-steps: \textbf{Node loading}, \textbf{Node placing}, and \textbf{Queue update}.

\subsubsection{Node loading}
The purpose of node loading is to determine which nodes and their versions will be added to the node tree.
As described in the background, a node and all its peer dependencies need to be visible to its parent node.
Therefore, the node loading process needs to load not only the current target node but also all its peer dependencies.

A necessary condition for the emergence of \PeerSpin is the occurrence of dependency version conflicts, which result in the replacement of one version with another.
During node loading, two types of conflicts may occur.
The peer dependencies in the set can conflict either with each other or with a common direct dependency of the parent node.
However, the dependency resolution algorithm does not terminate due to these conflicts, but instead marks the conflicting dependencies as invalid and handles them in the subsequent node placing process.
These illegitimate edges can affect the state of the node tree during node placing, potentially causing the resolution algorithm to fall into an infinite loop and lead to \PeerSpin.

\begin{boxA}
\label{Insight1}
    \textbf{Insight-1}: Two types of conflicts (\textbf{\textit{Peer-to-Regular}} and \textbf{\textit{Peer-to-Peer}}) may occur during node loading. By default, the resolution algorithm takes a loose approach to avoid conflicts involving peer dependencies, thereby leaving the risk of \PeerSpin to subsequent processes. 
\end{boxA}

\begin{figure}[t]
    \centering
    \begin{subfigure}{0.4\linewidth}
        \centering
        \includegraphics[width=\linewidth]{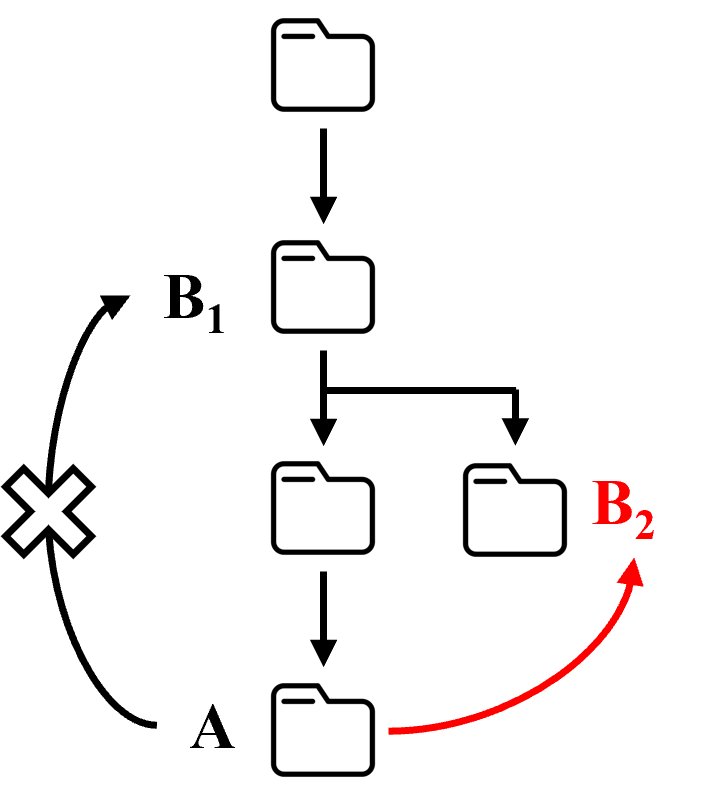}
        \caption{Node adding.}
        \label{fig:Node-Add}
    \end{subfigure}
    \begin{subfigure}{0.345\linewidth}
        \centering
        \includegraphics[width=\linewidth]{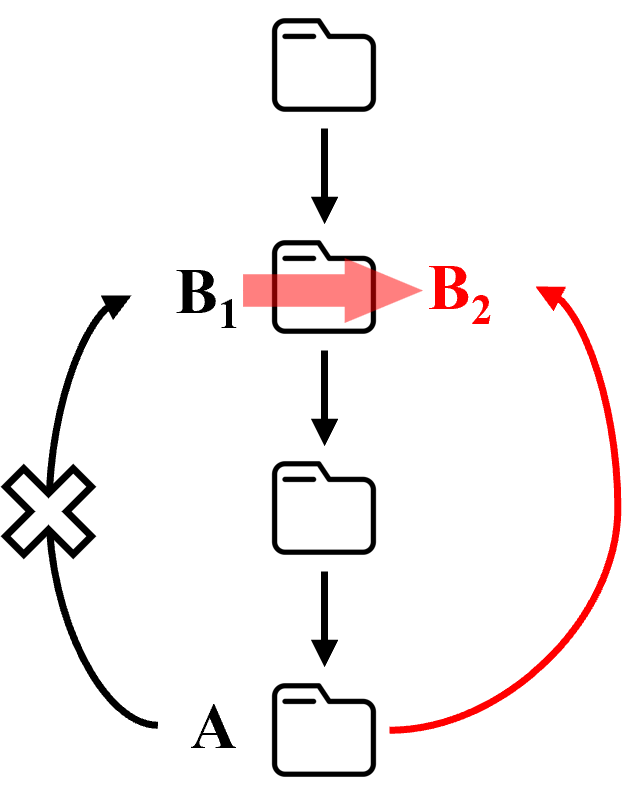}
        \caption{Node replacing.}
        \label{fig:Node-Replace}
    \end{subfigure}
    \vspace{5pt}
    \caption{The state of the node tree changes due to the add or replace node.}
    \label{fig:NodeTreeChange}
\end{figure}

\subsubsection{Node placing}
Adding a node and its PeerSet to a node tree as children of another node is called node placing.
According to the rules mentioned in \cref{DependencyLoading}, a node is more likely to be reused if it is placed at a shallow depth in the node tree.
Therefore, the core algorithm for node placement is to place the node as shallowly as possible in the node tree while ensuring it can be found by its PeerSource.

When placing a node in the node tree, depending on whether a child node with the same name as the node to be placed already exists at the intended location, it can be classified into the following four scenarios:
\begin{itemize}
    \item \textbf{ADD}: The current location doesn't have a child node with the same name as the node to be placed, so the node can be placed directly at the current position.
    \item \textbf{KEEP}: The current location already has a child node with the same name as the node to be placed, and its version satisfies the dependency requirements.
    \item \textbf{REPLACE}: The current location has a child node with the same name but a different version than the node to be placed, so the new node replaces the existing one.
    \item \textbf{CONFLICT}: Placing a node at this position disrupts a currently satisfied dependency in the node tree.
\end{itemize}

\normalem
\begin{algorithm}[t]
\caption{Node Placing.}
\label{algorithm:NodePlacing}
\kwInit{$ n,\ the\ node\ to\ be\ placed $}
$ pos \gets getStartPos() $ \\
$ type \gets CONFLICT $ \\
\While{$ n $} {
    $ t \gets canPlace(n, pos) $ \\
    \If {$ t == CONFLICT $ } {
        $ break $ \\
    }
    $ type \gets t$ \\ 
    $ lastPos \gets pos $ \\
    $ pos \gets pos.upDirectory() $ \\
}

\eIf{$ type == CONFLICT $}{ 
    $ exit() $ \\
}{\If{$ type == KEEP $}{ 
      $ return $ \\
    }
}

\eIf{$ type == REPLACE$}{
    $ lastPos.replace(n) $ \\
}{$ lastPos.place(n) $ \Comment*[r]{$ type == ADD $}} 

$ TreePrune() $
\end{algorithm}

In the four scenarios above, the KEEP type will not change the status of the node tree.
If CONFLICT occurs, the node placement process will continue searching for another position. 
However, ADD and REPLACE, may change the status of the node tree, potentially invalidating valid edges.
These invalid edges need to be reprocessed later to restore their validity.
For example, in \cref{fig:NodeTreeChange}, introducing $B_2$ deeper into the tree than $B_1$, or replacing $B_1$ with $B_2$, will cause $A$ to fail to load the correct dependency version according to npm's dependency loading rules.

The pseudo-code for the node placing algorithm is shown in \cref{algorithm:NodePlacing}.
The node placement process begins at the deepest possible position where the node can be placed.
For peer dependencies, this is at the same level as the PeerEntry.
The process then iterates upward through the directory levels of the node tree, finding the last position where no CONFLICT occurs (lines 3\textasciitilde10).
If no suitable position is found, the resolution algorithm reports an error and terminates (line 13).
The algorithm does nothing if the placing type is KEEP (line 15\textasciitilde17).
In other cases, the node is replaced or added at \code{lastPos} (lines 19\textasciitilde22).
Finally, any nodes in the node tree affected by the ADD or REPLACE must be deleted (line 24).

Taking \cref{fig:PeerSpin} as an example, when node \code{antd} is placed, its PeerSet, which includes \code{react-dom} and \code{react}, is also placed.
Since no package with the same name exists in the installation directory, \code{react} is placed in the root directory.
The algorithm then attempts to place \code{draft.js} and its PeerSet, which includes incompatible versions of \code{react}.
Node replacing is performed. After the replacement, since \code{react-dom}'s dependency on \code{react} is not satisfied, \code{react-dom} is removed. 
Additionally, \code{antd}'s dependency on \code{react-dom} is marked invalid.
During the subsequent reprocessing of \code{react-dom}, a similar replacement occurs, causing the resolution algorithm to enter an infinite loop.

\begin{boxA}
    \textbf{Insight-2}: During the node placement process, the status of the node tree may change as nodes are added or replaced. However, since the algorithm relies solely on the current status of the node tree, the process of changing the node tree can enter a loop, leading to \PeerSpin.
\end{boxA}

\subsubsection{Queue updating}
The npm dependency resolution relies on a breadth-first traversal algorithm that uses a queue of nodes to manage the processing and determines termination based on whether the queue is empty.
Therefore, the queue must be properly updated.
The queue update process involves adding two main types of nodes: those placed in the node tree during the current round of node placement, and those in the tree with invalidated edges due to node additions or replacements.
In the subsequent out-queue processing, the edges of these nodes are resolved, and new nodes are loaded and placed in the node tree to validate these edges.
If a sequence of nodes continuously cycles in and out of the queue as they are processed by the algorithm, generating an infinite loop of enqueue and dequeue operations, the queue will never be empty. Consequently, the algorithm will get stuck in repeated processing of that sequence of nodes, resulting in \PeerSpin.

If the nodes and their dependency satisfaction remain constant in the node tree, they should not re-enter the queue.
However, \PeerSpin involves the same nodes being enqueued repeatedly, and the edges of these nodes being processed repeatedly.
This means that the status of these edges must change from \textit{valid} to \textit{invalid} during node placing.
Therefore, between two instances of the same node being enqueued, there must be a replacement of either that node or its dependencies.
Replacing the node itself may invalidate the edge pointing to this node, causing the source node of the edge to be added back to the queue. This node is then re-placed and re-queued.
If a dependency of this node is replaced, the edge pointing to this dependency may become invalid, causing this node to be added back to the queue and its dependency to be placed again.
Additionally, to create a looping replacement, the node replacements leading to the loop must be bidirectional. This means that the replacing node and the replaced node must interchangeably replace each other during their respective node placement processes.
Therefore, these two nodes must have the same name, indicating that they represent the same package.

\begin{boxA}
    \textbf{Insight-3}: Two nodes responsible for \PeerSpin must be peer dependency nodes with the same name but different versions. As these two nodes replace each other, their associated dependencies become unsatisfied, causing the nodes to be re-queued and waiting to be placed again, thus creating a cyclic sequence.
\end{boxA}

\subsection{RQ2: Dependency Patterns}
\label{Patterns}

\begin{figure}[t]
    \centering
    \begin{subfigure}{0.43\linewidth}
        \centering
        \includegraphics[width=0.8\linewidth]{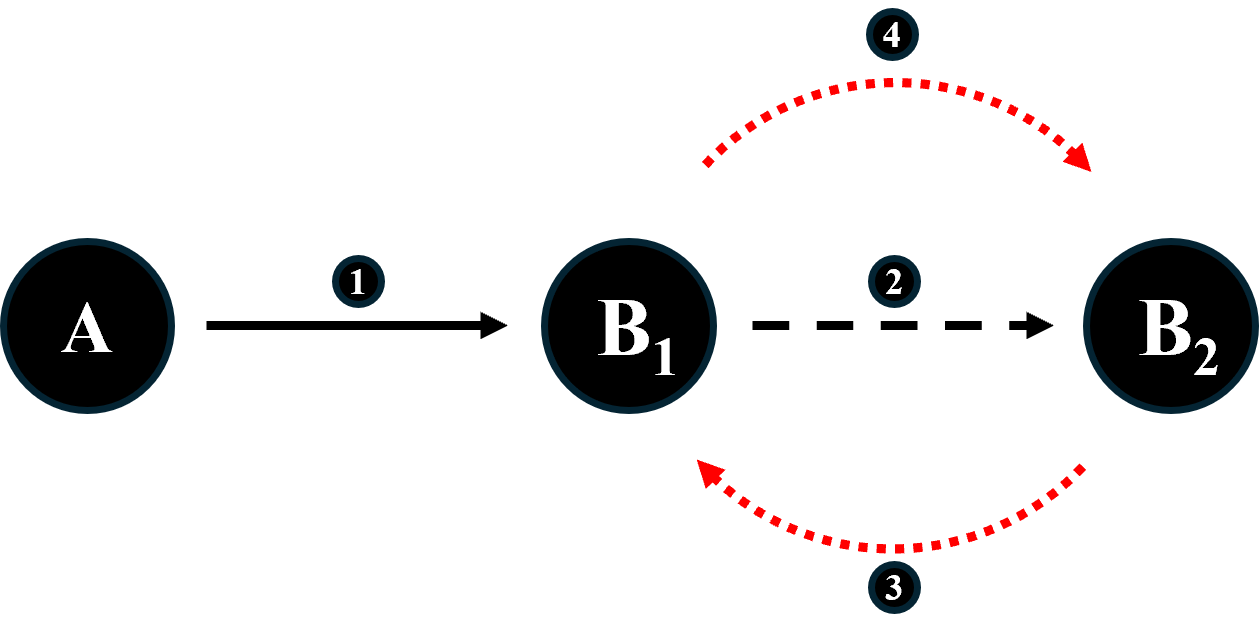}
        \caption{Pattern A}
        \label{fig:pattern1}
    \end{subfigure}
    \begin{subfigure}{0.55\linewidth}
        \centering
        \includegraphics[width=0.95\linewidth]{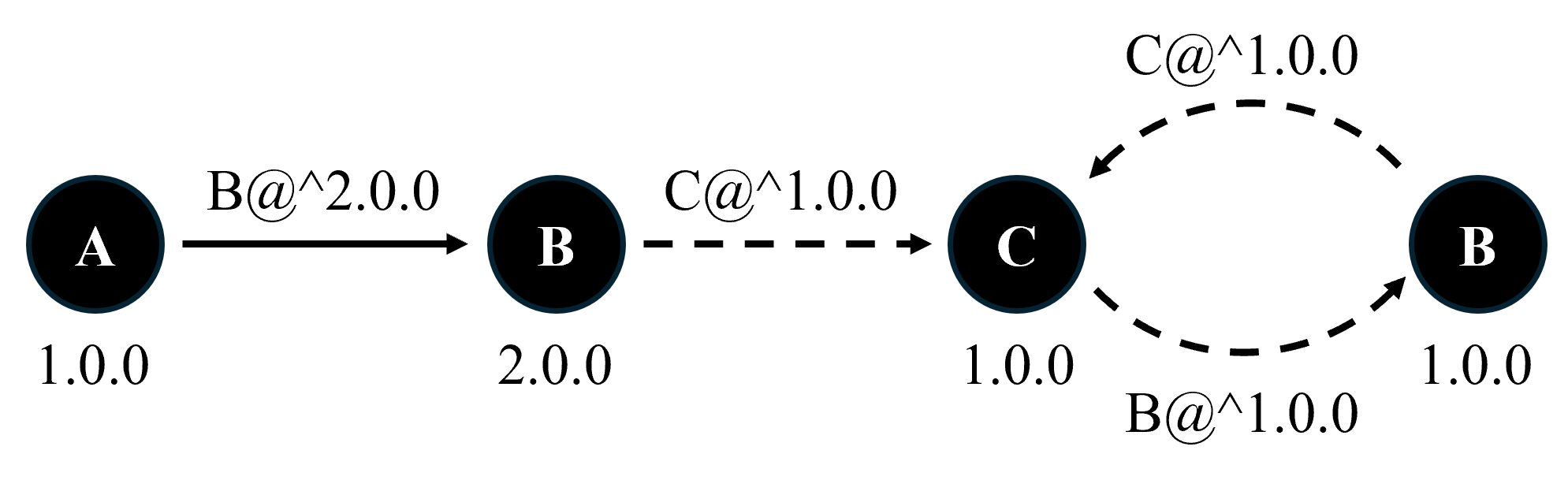}
        \caption{Pattern A example}
        \label{fig:pattern1-example}
    \end{subfigure}
    \vfill
    \begin{subfigure}{0.43\linewidth}
        \centering
        \includegraphics[width=0.8\linewidth]{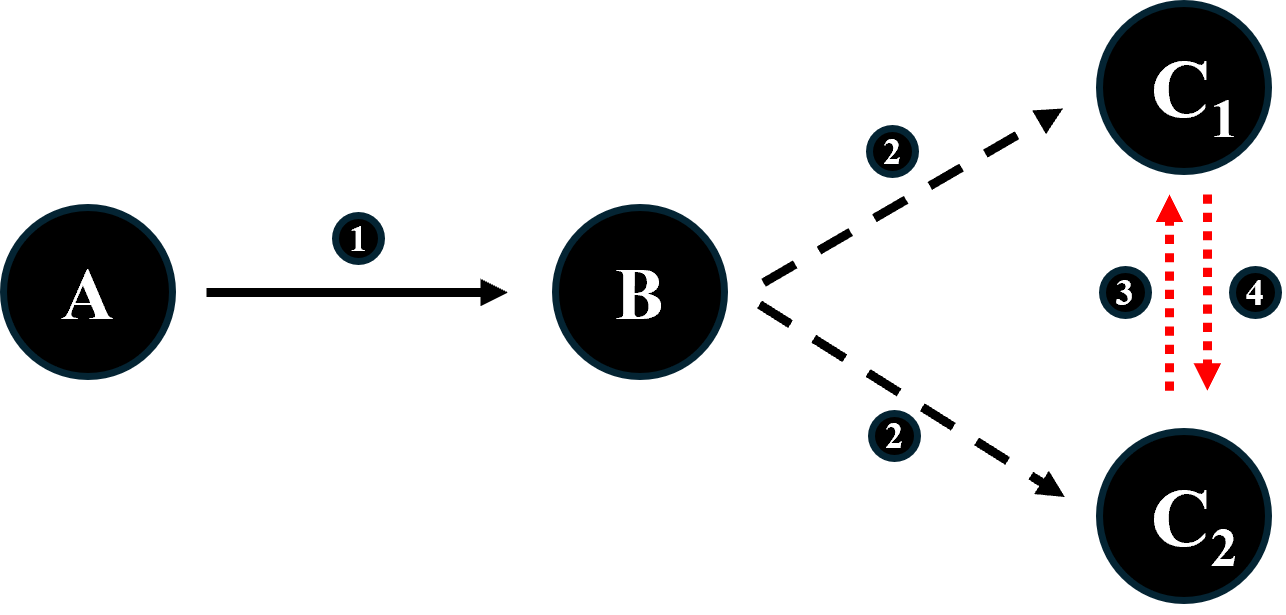}
        \caption{Pattern B}
        \label{fig:pattern2}
    \end{subfigure}
    \begin{subfigure}{0.55\linewidth}
        \centering
        \includegraphics[width=0.95\linewidth]{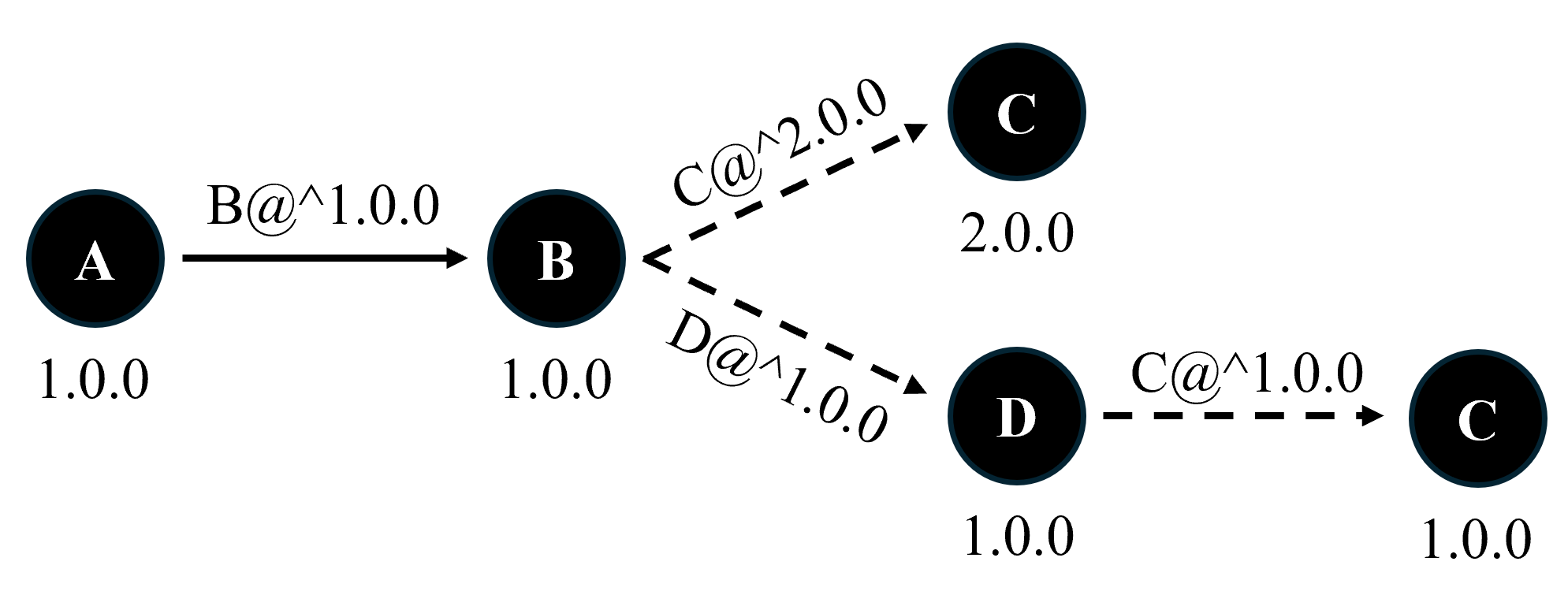}
        \caption{Pattern B example}
        \label{fig:pattern2-example}
    \end{subfigure}
     \caption{Two patterns and examples for peer dependency resolving loop. $\rightarrow$ means regular dependency. $\dashrightarrow$ means peer dependency. {\color{red}{$\dashrightarrow$}} means node replacement.}
\end{figure}

Insights 1 to 3 reveal that two types of conflicts (\textbf{\textit{Peer-to-Regular}} and \textbf{\textit{Peer-to-Peer}}) are not adequately managed during node loading. These conflicts result in node replacement during the subsequent node placing process, invalidating previously satisfied dependencies in the node tree and causing repeated entries of related nodes into the queue, leading to \PeerSpin.
Based on these insights, we categorize two peer dependency patterns that can lead to \PeerSpin.

\vspace{3pt}
\begin{boxA}
    \noindent\textbf{Pattern A}: Peer dependencies conflict with the common dependencies of their PeerSource. This conflict results in node replacements, causing the dependency of PeerSource on PeerEntry to fail. Consequently, PeerSource repeatedly enters and exits the queue, and the conflicting dependency is processed again, leading to a \PeerSpin.
\end{boxA}
%
In \cref{fig:pattern1}, \ding{182} $A$ directly has a regular dependency $B_{1}$, \ding{183} $B_{1}$ indirectly has a peer dependency, which points to itself but with a different version $B_{2}$. Since the two versions are incompatible, \ding{184} \ding{185}, they will be stuck in a loop of replacing each other.

For example, in \cref{fig:pattern1-example}, \code{A@1.0.0} has a regular dependency \code{B@2.0.0} while \code{B@2.0.0} depends on \code{B@1.0.0} through \code{C@1.0.0} as peer dependency. 
When \code{B@2.0.0} is out of the queue to be processed, the node loading process will load PeerSet of \code{B@2.0.0}, which are \code{B@2.0.0} and \code{C@1.0.0} (\code{B@1.0.0} can not be loaded due to \code{B@1.0.0} already existing).
After \code{B@2.0.0} and \code{C@1.0.0} are placed in the node tree, the algorithm needs to handle the dependency of \code{C@1.0.0} on \code{B@1.0.0}. 
It will replace \code{B@2.0.0} to \code{B@1.0.0} to make the dependency of \code{C@1.0.0} on \code{B} valid, but it simultaneously makes the dependency of \code{A@1.0.0} on \code{B} invalid, causing the PeerSource \code{A@1.0.0} to enter the queue again. Then \code{A@1.0.0} will be processed again and replaced with \code{B@1.0.0} to \code{B@2.0.0}, which makes the dependency of  \code{C@1.0.0} on \code{B} invalid again and \code{C@1.0.0} into the queue again. The algorithm falls into an infinite loop.

\begin{boxA}
    \noindent\textbf{Pattern B}: Conflicts between peer dependencies within a PeerSet result in node replacements, causing the dependency of PeerEntry on the node within the PeerSet to fail. As a result, PeerEntry repeatedly enters and exits the queue, leading to the conflicting dependency being processed again, which in turn leads to a \PeerSpin.
    
\end{boxA}

In \cref{fig:pattern2}, \ding{182} $A$ has a regular dependency $B$, \ding{183} $B$ has two peer dependencies on different dependency paths pointing to two incompatible versions of $C$. \ding{184} \ding{185} These two version of $C$ will fall into a loop of replacing each other.

In \cref{fig:pattern2-example}, \code{A@1.0.0} regular dependency on \code{B@1.0.0}, 
\code{B@1.0.0} has two peer dependencies on \code{C@2.0.0} and \code{D@1.0.0}.
When \code{B@1.0.0} is out of the queue to be processed, the node loading process will load PeerSet of \code{B@1.0.0}, which are \code{B@1.0.0}, \code{C@2.0.0}, and \code{D@1.0.0} (\code{C@1.0.0} can not be loaded due to \code{C@2.0.0} already existing).
After the nodes of PeerSet are placed in the node tree, the algorithm needs to handle the dependency of \code{D@1.0.0} on \code{C@1.0.0}. 
It will replace \code{C@2.0.0} to \code{C@1.0.0} to make the dependency of \code{D@1.0.0} on \code{C} valid, but it simultaneously makes the dependency of \code{B@1.0.0} on \code{C} invalid, causing the PeerEntry \code{B@1.0.0} to enter the queue again. 
Then \code{B@1.0.0} will be processed again and replaced \code{C@1.0.0} with \code{C@2.0.0}, which makes the dependency of \code{D@1.0.0} on \code{C} invalid again and \code{D@1.0.0} into the queue again. The algorithm falls into an infinite loop.

Notably, the patterns include only the minimal node sets required to produce \PeerSpin.
These patterns are not mutually exclusive but have the potential to occur simultaneously, and multiple times in one dependency graph.
Even when adding intermediate nodes to the dependency graph, as long as no conflicts are introduced and the type of dependency of the inserted node matches the type of the inserted edge, the \PeerSpin problem remains.
That also brings the challenge of \PeerSpin detection.
\section{PeerSpin Detection}
\label{Diagnosis}
\begin{figure}[t!]
    \centering
    \includegraphics[width=\linewidth]{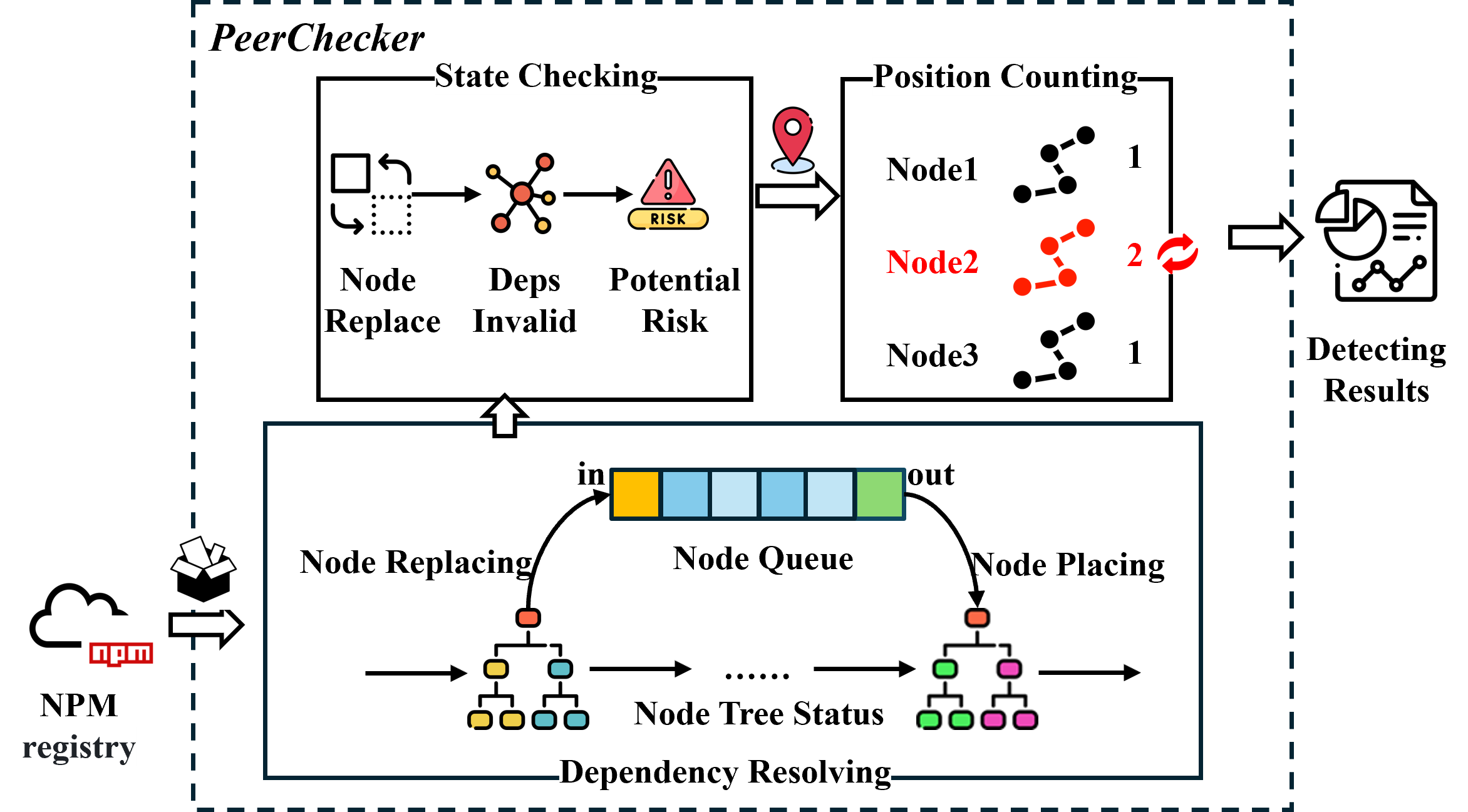}
    \caption{Overview of \textit{\PeerChecker}.}
    \label{fig:PeerChecker}
\end{figure}

To quantify the impact of \PeerSpin on the entire npm ecosystem, we need to detect it accurately and efficiently.
In this section, we introduce \PeerChecker, a tool designed to investigate the impacts of \PeerSpin in the npm ecosystem.
~\Cref{fig:PeerChecker} provides an overview of \PeerChecker. \PeerChecker intercepts the state of the node tree, performs status checks, and records the position of the node to detect \PeerSpin.

\subsection{Design}
In RQ1, we find that the direct cause of \PeerSpin is the node replacement cycle. This cycle arises when the replacement of nodes leads to dependency conflicts, causing some nodes to re-enter the queue repeatedly.
Therefore, we propose a novel detection technique, \Tech, that identifies nodes that may trigger \PeerSpin by detecting dependency version conflicts due to node replacement.

The basic idea of this detection approach is to mark nodes that may cause subsequent loops based on the current state of the node tree after node replacement and to monitor the behavior of these nodes during subsequent dependency resolution.

First, we monitor the behavior of each node replacement during dependency resolution, intercepting and collecting the current state of the node tree.
After a node in the tree is replaced, the dependency satisfaction of its PeerSource and PeerEntry may be broken (as shown in the examples of pattern A and pattern B).
This causes the PeerSource or PeerEntry to re-enter the queue.
Subsequent PeerSource and PeerEntry being processed again may result in node placement of another version of the target node, leading to node replacement.
Thus, we consider the replaced nodes, which cause the dependency satisfaction of PeerSource or PeerEntry to break, as \textit{risky nodes}. 

Second, for each node in the node tree, its position determines how it is found according to the dependency loading rules. Therefore, the position information must be unique.
We record and count the position information for all \textit{risky nodes}.
If a node is replaced multiple times at the same location, it indicates that the node has generated a loop replacement.

\normalem
\begin{algorithm}[t]
\caption{\PeerSpin Detection.}
\label{alg:Detection}
\kwInit{$ N^{'},\ node.$}
\kwInit{$ T,\ node\ tree.$}
$ T.replace(N, N^{'}) $ \\
$ Risky \gets Flase $ \\
$ Source \gets N.getSource() $ \\ 
$ Entry \gets N.geEntry() $ \\
\For {$ edge \in Source.edges() $} {
    \If{$ edge.to == Entry  \And \neg edge.isValid()$} {
        $ PotentialRisk \gets True $ \\
        $ goto\ PC $ \\
    } 
} 
\For {$ edge \in Entry.edges() $} {
    \If{$ edge.to == N^{'} \And \neg edge.isValid() $} {
        $ Risky \gets True $ \\
        $ goto\ PC $ \\
    } 
} 
$ PC: $ \\
\If{ $ Risky $ } {
    $ pos \gets getPosition(N^{'}, T) $ \\
    $ PosCount[(N^{'}, pos)] += 1 $ \\
    \If {$ PosCount[(N^{'}, pos)] > 1 $} {
        $ return\ \PeerSpin $
    }
}
\end{algorithm}

\Cref{alg:Detection} presents the pseudo-code of the detection algorithm. When node replacement occurs, we calculate the PeerSource and PeerEntry (lines 1-4).
Then we first check the dependency edge of \code{PeerSource} on \code{PeerEntry}. If the edge is invalid, it indicates that the dependency satisfaction is broken (lines 5 and 6).
\code{PeerSource} needs to re-enter the queue which may result in $N^{'}$ being replaced by $N$ in subsequent processing. 
If \code{PeerSource} is not affected, we proceed to check \code{PeerEntry}. 
If the dependency of \code{PeerEntry} on the replaced node ($N^{'}$) is broken, \code{PeerEntry} needs to re-enter the queue.
In both scenarios, we need to mark node $N^{'}$ as a \textit{risky node} (lines 7 and 13).
If there is a risk node, the algorithm performs position counting (\code{PC}) starting at line 18.
We use the path from the root node of the node tree to the target node as the location information and use the pair of the node and its position as the key to count the occurrences.

Take \cref{fig:pattern1-example} as an example: when \code{B@1.0.0} replaces \code{B@2.0.0}, the detection algorithm first checks the \textit{PeerSource} of \code{B}.
The algorithm recognizes that this node replacement makes the regular dependency from \textit{PeerSource}, \code{A@1.0.0}, on \code{B} invalid.
It then marks \code{B@1.0.0} as a risky node and the position is counted.
In subsequent dependency resolution, \code{B@1.0.0} and \code{B@2.0.0} continue to substitute for each other and repeat.
The second time \code{B@1.0.0} replaces \code{B@2.0.0}, the detection algorithm performs similar checks as above, then the position of \code{B@1.0.0} is counted again,   
The algorithm detects that the same node is placed in the same position multiple times and reports \PeerSpin.

\subsection{Implementation}
Leveraging the \Tech technique, we developed a tool called \PeerChecker to detect issues across the entire ecosystem.
\Cref{fig:PeerChecker} provides an overview of \PeerChecker. 
The major challenge is scaling this detection approach to the vast number of packages in the npm ecosystem.
The npm client is implemented entirely in JavaScript, and the database backend it uses is not a high-speed database but often slower with lower throughput.
These factors contribute to reducing the speed of dependency resolution for npm clients.
As a result, we cannot use the npm client to conduct dependency analysis and detect \PeerSpin at the ecosystem scale.
Additionally, existing dependency resolution tools don’t consider peer dependencies, so we can’t reuse them.
To overcome this challenge, \PeerChecker implements an efficient dependency resolution algorithm that simulates the node loading, node placing, and queue update processes.
During the simulation, \PeerChecker monitors the state of the node tree, performs status checks, and counts positions to detect \PeerSpin.
To achieve higher performance, \PeerChecker is implemented entirely in \code{C++}, with data structures and algorithm steps optimized for accuracy and adapted to faster storage databases (e.g., Redis\cite{Redis}) and data formats.

\begin{table}
\centering
\renewcommand{\arraystretch}{1.5}
\caption{Detection accuracy.}
\resizebox{.8\linewidth}{!}{%
\begin{tabular}{ccc}
\hline
\backslashbox[45mm]{\textbf{npm client}}{\textbf{\PeerChecker}} & \textbf{Positive} & \textbf{Negative} \\ \hline
\textbf{\PeerSpin} & 7,300 & 0 \\
\textbf{Non-\PeerSpin} & 0 & 73,000 \\ \hline
\end{tabular}%
}
\label{tab:detectionAccuracy}
\end{table}

\subsection{Evaluation}
Our experiments run on Ubuntu 22.04 LTS with AMD EPYC 9654 96-Core Processor @3.55 GHz. The npm client version is 10.2.4.
To ensure that the meta-information used for dependency resolving is consistent and avoids network latency, both the npm client and \PeerChecker use a local mirror database in this test.

\subsubsection{Detection accuracy}
To assess the accuracy of PeerSpin detection, we used the npm client to verify the results obtained by \PeerChecker.
The test dataset sampling rules are as follows:

The testing dataset consists of \textit{Positive} and \textit{Negative} samples.
\begin{itemize}
    \item \textbf{\textit{Positive}}. We randomly selected 10\% of the package versions detected by \PeerChecker as \PeerSpin, 7,300 package versions in total.

    \vspace{5pt}
    \item \textbf{\textit{Negative}}. We denote the number of \PeerSpin package versions detected by \PeerChecker as $x$ and the number not detected as $y$. 
    The rate of occurrence of \PeerSpin is $P = y/(x+y)$.
    We then randomly selected undetected package versions according to this occurrence rate, resulting in a total of 73,000 package versions.
\end{itemize} 

We used the npm client to install the sampled package versions in an empty directory and developed an automated installation log analyzer.
\PeerSpin was considered to have occurred if the analyzer detected a loop log sequence created by node placement in the log.

As shown in \cref{tab:detectionAccuracy}, for all \textit{Positive} samples, the npm client occurs \PeerSpin. For all \textit{Negative} samples, the npm client completes the installation successfully.
The accuracy of \PeerChecker, designed based on two basic patterns, is confirmed.

\subsubsection{Performance}
The performance evaluation focuses on the time efficiency of the dependency analysis and \PeerSpin detection tools discussed in this paper.
The tests were performed on a random sample of all package versions, with a sample size of 1,000 versions.

For testing, \PeerChecker was run in daemon mode, using a local cached database as a data source in the data collection section of npm, with the thread pool size set to 16.
For comparison, the npm client also uses the same local cached database and installs the same package versions in \code{dry-run} mode. It only runs the resolution algorithm without performing any real dependency installation.
It also maintained 16 npm client processes in parallel.

The results of the performance tests are presented in \cref{tab:performance}
\PeerChecker is approximately 14 times faster than the npm client at resolving dependencies, even though the npm client does not perform any real dependency installation.
When performing dependency analysis on over 36 million package versions in the entire npm open-source ecosystem, \PeerChecker can reduce the overall time consumed from approximately one month with the npm client to about three days, which is acceptable.
\section{Ecosystem-scale Study}
\label{Study}
We conducted a large-scale analysis to understand peer dependency usage and the impact of \PeerSpin on npm ecosystems.
Our source data is derived from the official npm registry snapshot taken on October 1, 2023. This dataset includes metadata and information on 36,767,234 package versions, including package names, version numbers, dependencies, and release dates.
To drive our study, we raise two research questions (RQs).

\begin{itemize}[]
\item \textbf{RQ3 (Usage).} How many npm packages use peer dependency?
\item \textbf{RQ4 (Impact).} How does \PeerSpin affect the npm ecosystem?
\end{itemize}

\subsection{RQ3: Usage of Peer Dependency}

\begin{table}[t]
\centering
\renewcommand{\arraystretch}{1.5}
\caption{Performance testing results.}
\resizebox{.95\linewidth}{!}{%
\begin{tabular}{cccc}
\hline
 & \textbf{Execution Time} & \textbf{Average Time} & \textbf{Speed Ratio} \\ \hline
\textbf{npm client} & 1m19.142s & 79.142ms & 1x \\
\textbf{PeerChecker} & 0m5.573s & 5.573ms & 14.201x \\ \hline
\end{tabular}%
}
\label{tab:performance}
\end{table} 

\begin{figure}
    \centering
    \includegraphics[width=\linewidth]{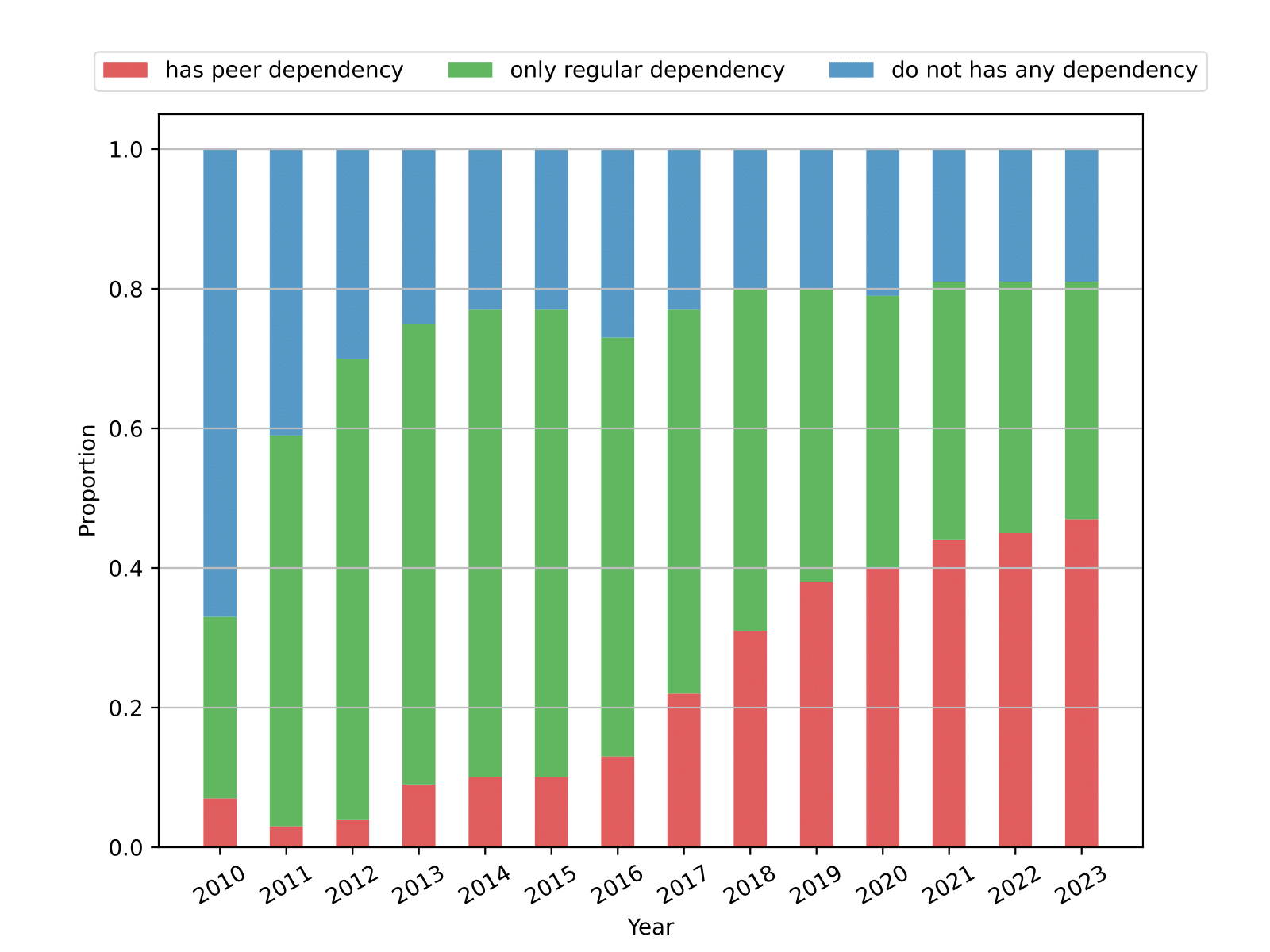}
    \caption{Proportion of dependency by type.}
    \label{fig:TypeEvolution}
\end{figure}

To analyze the scale and preference of peer dependencies in the npm ecosystem, we examine peer dependencies from two perspectives: how many packages use peer dependencies, and how many are used as peer dependencies?

Peer dependencies are prevalent in the npm open-source ecosystem. 
Approximately 61.15\% of packages used peer dependencies.
Additionally, to further analyze the use trend of peer dependencies, we categorized new package versions each year by dependency type and tracked their proportions.
There are three categories: no dependencies, only regular dependencies, and peer dependencies.
The results are presented in ~\cref{fig:TypeEvolution}. 
The proportion of new package versions with peer dependencies (39.89\%) added each year surpassed those without peer dependencies (39.04\%) in 2020. This trend has continued to grow faster, reaching 45.66\% in 2023.
The temporal evolution of these proportions indicates that peer dependencies are gaining widespread popularity.

\begin{table}[t!]
\vspace{5pt}
\centering
\renewcommand{\arraystretch}{1.5}
\caption{The 5 most influential package versions.}
\resizebox{.7\linewidth}{!}{%
\begin{tabular}{ccc}
\hline
\textbf{Name/Version} & \textbf{\begin{tabular}[c]{@{}c@{}}\# Peer \\ Dependent\end{tabular}} & \textbf{\begin{tabular}[c]{@{}c@{}}\ Released \\ Year\end{tabular}} \\ \hline
react@18.2.0 & 2,482,920 & 2022 \\
supports-color@5.5.0 & 2,181,088 & 2018 \\
react@16.14.0 & 1,955,730 & 2020 \\
@babel/core@7.22.20 & 1,899,125 & 2023 \\
react-dom@18.2.0 & 1,665,575 & 2022 \\ \hline
\end{tabular}%
}
\label{tab:Top5}
\end{table}

Our analysis also found that only a very small subset of packages are used as peer dependencies.
In all the package versions that are used as dependencies in other packages, only 4.4\% are used as peer dependencies.
This suggests that the widespread use of peer dependencies makes the stability of this subset critical.
Since the influence of dependencies in the npm open-source ecosystem is highly concentrated, we identified the most influential package versions, as shown in ~\cref{tab:Top5}.
The two incompatible versions of \code{react} are heavily relied upon by downstream packages, which can result in the same downstream package directly or indirectly depending on the peer dependencies of both incompatible versions, as shown in ~\cref{fig:PeerSpin}. 

\subsection{RQ4: Impact of \PeerSpin}
We detect \PeerSpin issues for the entire npm ecosystem and find 72,968 versions of 5,662 packages that suffer from \PeerSpin. 
We provided detailed logs to package maintainers, encouraging them to update their dependency configurations and release new versions compatible with the affected versions.
We identified and reported 100 problematic packages to corresponding developers at the time of writing, receiving 28 confirmations.
Among these, 19 maintainers appreciated our reports and committed to adjusting their dependency configurations in future versions.
Additionally, 5 packages have been migrated to new projects, and we find that the new projects are unaffected by the \PeerSpin.
For the remaining four packages, maintainers suggested alternative approaches, such as switching to  \code{yarn}\cite{yarn} or reverting to an older version of the npm client as a temporary workaround.

\begin{figure}[t]
    \centering
    \includegraphics[width=\linewidth]{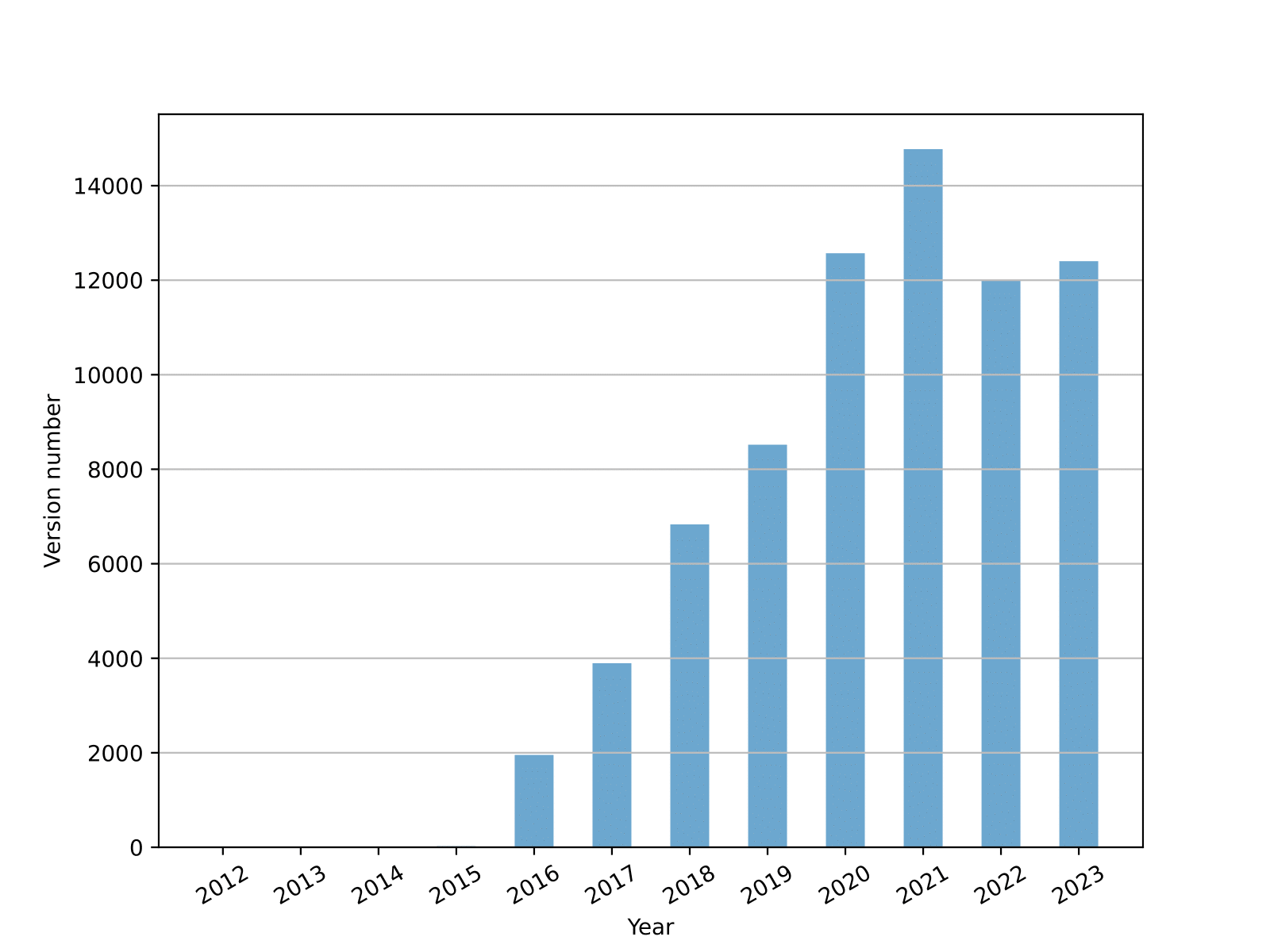}
    \caption{The number of problematic package versions of time evolution.}
    \label{fig:LoopVersionEvolution}
\end{figure}

\Cref{fig:LoopVersionEvolution} shows the number of package versions in each year that are currently affected by \PeerSpin. 
There are minimal effects for packages released in 2015 and earlier. 
Since 2016, the number of problematic releases has grown, growing rapidly at an average rate of 52.6\%, peaking in 2021 before dropping slightly and remaining high (over 10,000 package versions per year). 

The turnaround for the number of affected package versions occurred in 2016 and 2021. 
The npm client discontinued the automatic installation of peer dependencies in \code{v3}. This feature was reinstated in \code{v7}.
The first official versions of \code{npm@v3} and \code{npm@v7} are released in June 2015 and Oct 2020 respectively.
Due to the lag of new versions of the \code{npm} being used on a large scale, it can be assumed that the timing of the update and popularization of the npm client version coincides with the timing of the temporal evolution of \PeerSpin issues.
Before \code{npm@v3}, the number of problems with released package versions was minimal due to the simplicity of applying peer dependencies and the small size of the npm ecosystem. 
Since \code{npm@v3} stopped the automatic installation of peer dependencies, peer dependency installation relied more on manual installation by developers, and the number of \PeerSpin issues increased rapidly along with the size of the npm ecosystem. 
\code{npm@v7} resumed the automatic installation of peer dependencies, which led to a more standardized installation of peer dependencies in the following years. \PeerSpin remains an issue due to historical legacy issues. 
In addition, the increasing complexity of dependencies due to the rapid development of the npm ecosystem is one of the main reasons for the increasing number of problematic versions, as the main factor in \PeerSpin is the conflicting versions of dependencies.

\begin{table}[t]
\vspace{5pt}
\centering
\renewcommand{\arraystretch}{1.5}
\caption{Top 10 critical packages that caused \PeerSpin.}
\resizebox{.95\linewidth}{!}{%
\begin{tabular}{cccc}
\hline
\textbf{Pkg} & \textbf{\begin{tabular}[c]{@{}c@{}}\# Weekly \\ Downloads\end{tabular}} & \textbf{\begin{tabular}[c]{@{}c@{}}\# Affected \\ Pkgs\end{tabular}} & \textbf{\begin{tabular}[c]{@{}c@{}}\# Affected \\ Vers\end{tabular}} \\ \hline
react & 23M & 2,833 & 41,758 \\
@angular/core & 3M & 490 & 4759 \\
eslint-plugin-flow type & 4M & 273 & 2558 \\
typescript & 55M & 267 & 869 \\
rxjs & 41M & 245 & 1362 \\
webpack & 25M & 140 & 1282 \\
react-dom & 21M & 121 & 2840 \\
eslint & 38M & 81 & 674 \\
@angular/compiler & 3M & 79 & 251 \\
graphql & 12M & 71 & 1556 \\ \hline
\end{tabular}%
}
\label{tab:top10}
\end{table}

The average dependency update lag in the npm ecosystem is 7 to 9 months~\cite{decan2018evolution}.
To see the potential risk posed by the dependency update lag, we analyze the time gap between the problematic version and the latest version of these packages.
Among these affected versions, approximately 8.7\% are the latest versions of their packages, and 52.26\% of these versions have a gap of less than 300 days.
This suggests that prolonged non-updating results in a certain number of packages not being able to install the latest version as well as more than half of the versions being in the update lag window, posing a threat to downstream packages and npm clients.

We have also analyzed the reasons why these packages appear \PeerSpin, the critical packages are shown in \cref{tab:top10}.
In the RQ3, we analyzed the dependency relationships within the \code{react} package. Our findings indicated that \code{react} has the highest number of peer dependents, which is reflected in the fact that the packages affected by \code{react} are also the most numerous.
The main reason these packages cause a looping problem is updates to their major version numbers. Suppose a package has multiple requirements for its dependency packages with different semantic version ranges when the dependency packages are updated. In that case, the npm client is more likely to have conflicting version selections for the dependency packages, which can result in two patterns described in RQ2, leading to problems.

\section{Limitation}
\label{Limitation}

\PeerChecker detects \PeerSpin based on dependency resolution and has promising results, but it still faces the following limitations.
 
First, while the local environment can affect the result of dependency resolution, \PeerChecker cannot collect relevant data from developers due to privacy concerns.
Therefore, \PeerChecker assumes dependency resolution from scratch. The process constructs the node tree and node graph independently of the local environment.
\PeerChecker also assumes the resolution does not use files like \code{package-lock.json}.
This means the process includes dependency resolution and version selection as usual rather than directly using the dependency directory tree specified in these files.
However, since local environments can complicate dependency resolution, we recommend that developers maintain an environment as clean as possible.

Second, the parameters for dependency resolution are set using the default parameters of the npm client during dependency installation. 
This means that parameters such as \code{global}, \code{workspace}, \code{prefer-dedupe}, \code{omit}, or any others that could change the resolution logic, dependency type, or installation result are neither used nor modified.
These operations may walk around the \PeerSpin issue but may introduce another dependency issue, i.e., dependency incompatibility, semantic inconsistencies, runtime errors, or other consequences in npm projects.

\section{Discussion}
\label{Discussion}
We discuss how to reduce the occurrence and impact of \PeerSpin from two perspectives: npm maintainers and developers.

\textbf{Maintainers of npm}. 
The npm client allows the installation of different package versions in separate subdirectories to address dependency conflicts.
However, this approach fails to address \PeerSpin, caused by conflicts in peer dependencies.
To avoid \PeerSpin, npm clients should audit peer dependencies and explicitly warn developers of version conflicts before installing.
Additionally, we suggest that the npm client implements a maximum iteration limit during peer dependency resolution, akin to the \code{ResolutionTooDeep}\cite{ResolutionTooDeep} exceptions in Pip/Python, to prevent infinite loops.
For scenarios involving incompatible peer dependencies, we suggest the npm client adopt a strategy similar to \textit{name mangling}\cite{NameMangling} in Cargo/Rust, which distinguishes multiple dependency versions by unique names and version numbers to promote greater flexibility and system stability.

\textbf{Developers}.
For developers, when declaring peer dependencies, it is important to be aware of the potential \PeerSpin with other dependencies. 
In addition, it is not recommended that developers use fixed versions of peer dependencies but instead use loosely-range versions. 
Due to the dynamic of the npm ecosystem, developers should check for peer dependency compatibility regularly. 
Developers should keep not only peer dependencies but also regular dependencies as simple as possible. 
The one reason for \PeerSpin arises is the complex dependency relationship. Removing the redundant dependencies can facilitate the resolving process and reduce the risk of compatibility issues. 
To further assist developers, our tool, \PeerChecker can help identify the presence of \PeerSpin in peer dependency configurations.
Additionally, by analyzing detailed dependency chains, \PeerChecker identifies which peer dependencies require modification.
Using this information, developers can resolve \PeerSpin by adjusting conflicting version limits or migrating problematic peer dependencies into regular dependencies, ensuring successful dependency installation.

\section{Related Works}
\label{RELATEDWORK}
\noindent\textbf{Dependency Analysis}.
Dependencies are fundamental to modern software development and have received significant research attention.
Latendresse et al.~\cite{latendresse2022not} investigate the impact of production-released project dependencies on security and management. 
Abdalka-Reem et al.~\cite{abdalkareem2017developers, abdalkareem2017reasons} analyze trivial dependency packages in npm and JavaScript applications.
Decan and Mens~\cite{decan2019package} focus on the usage and compatibility issues of semantic versioning. 
Wang et al.\cite{wang2018dependency, wang2020watchman, wang2021hero} and Li et al.~\cite{li2022nufix} studied dependency conflicts across ecosystems and developed detection tools, though similar research in the npm ecosystem is lacking.
Wittern et al.~\cite{wittern2016look} performed the first large-scale analysis of the npm ecosystem, revealing its evolution through direct dependencies. 
Decan et al.~\cite{10.1145/2993412.3003382, decan2017empirical, decan2019empirical}  compared dependency graph evolution in various programming ecosystems. 
Jens et al.~\cite{dietrich2019dependency} reviewed dependency declarations across 17 ecosystems and categorized version classifications.
However, most previous studies focus only on regular dependencies, overlooking the different types of dependencies that have different implications for projects and ecosystems. 
In contrast, our work is the first to uncover a critical issue, \PeerSpin, in the npm ecosystem arising specifically from peer dependencies, a previously unstudied dependency type.

\vspace{5pt}
\noindent\textbf{Package Manager Analysis}. 
Different package managers have different dependency management strategies. Existing works focus on comparing dependency management strategies across different ecosystems.
Pietro et al.~\cite{9054837} systematically compared resolvers in various dimensions, including
conflict solutions, range modifiers, etc. 
Stringer et al.~\cite{9359290} analyzed 14 package managers and demonstrated that technical lag, which can lead to security vulnerabilities and make the software more vulnerable, is common.
Hanus et al.~\cite{hanus2018semantic} designed a semantic version checker for verifying version semantics in package managers and integrated it into the package manager to fully automate the checking process.
Jacobs et al.~\cite{jacobs2019comparison} analyzed the security properties of package managers and uncovered several design-level vulnerabilities.
Pereira et al.~\cite{pereira2021using} reviewed the use of package manager and repository metrics to assess the security of npm packages.
However, these works ignore the analysis of specific implementations of package manager policies.
Specific features and limitations of a package manager are often tied to its internal implementation.
Without a detailed examination, researchers may not fully understand why certain features are present or absent or why specific limitations or issues exist.
Our study provides a detailed analysis of how the package manager handles peer dependencies within the npm ecosystem. We believe this work offers valuable insights that can assist both developers and researchers in achieving a deeper understanding of npm dependency management.

\vspace{5pt}
\noindent\textbf{Dependency-Driven Vulnerability Propagation Analysis}.
The security implications of package dependencies, particularly the vulnerabilities they introduce, have been extensively studied in prior research.
Zerouali et al.~\cite{zerouali2018empirical}  analyzed version lag between package dependencies using the npm dataset from Libraries.io~\cite{LibIO}.
Chinthanet et al.~\cite{chinthanet2021lags} examined version lag and constraints on downstream packages.
Decan et al.~\cite{SecImpact} assessed the impact of 400 security vulnerabilities across 610,097 package versions.
Zimmermann et al.~\cite{zimmermann2019small} analyzed npm security risks through dependencies and vulnerabilities, using direct dependencies.
Zerouali et al.~\cite{zerouali2022impact} studied the impact of vulnerable dependencies on both npm and RubyGems ecosystems. 
Liu et al.~\cite{VulviaDepTree} proposed a knowledge graph-based dependency solution, studying the security threats posed by dependency tree vulnerabilities on a large scale.
Alfadel et al.~\cite{10.1145/3571848} examined the impact of vulnerabilities at various stages of disclosure.
Wang et al.~\cite{wang2023plumber} analyzed blocked updates in critical dependency chains.
However, these works predominantly address the security risks associated with vulnerable code introduced by third-party dependencies. 
In contrast, our study highlights a novel concern: even when the dependency code is non-vulnerable, the intricate relationships among dependencies can still result in severe issues, ultimately affecting the stability and security of the ecosystem.

\section{Conclusion}
\label{CONCLUSION}
In this paper, we conduct the first in-depth study to analyze the cause of \PeerSpin and its impacts on the npm ecosystem.
We are modeling the npm client dependency resolution process to uncover the root cause of \PeerSpin and two peer dependency patterns are also proposed to guide the following detection.
For \PeerSpin detection, we propose a novel technique, \Tech, for efficient and accurate detection of \PeerSpin. We also designed and implemented a tool called \PeerChecker to expand detection to ecosystem-scale.
We identified 72,968 package versions suffering from \PeerSpin issues. We report problems and get feedback. All experimental data in this paper are available at \url{https://github.com/ZJU-SEC/PeerChecker}.

\section{Acknowledgment}
The authors would like to thank all reviewers sincerely for their valuable comments. This work is partially supported by the
National Key R\&D Program of China (2022YFB3103900).

\bibliographystyle{unsrt}
\bibliography{ref}

\end{document}